# Neuropeptide Y is up-regulated and induces antinociception in cancer-induced bone pain

Marta Diaz-delCastillo[a], Søren H. Christiansen[b], Camilla K. Appel[a], Sarah Falk[a,b], David P. D. Woldbye[b], Anne-Marie Heegaard[a]

[a] Department of Drug Design and Pharmacology, Faculty of Health and Medical Sciences, University of Copenhagen. Jagtvej 160, Copenhagen Ø, DK-2100 (Denmark).

[b] Department of Neuroscience, Faculty of Health and Medical Sciences, University of Copenhagen. 3 Blegdamsvej, Copenhagen N, DK-2200 (Denmark).

## Abbreviations

| | |
|---|---|
| BSA | Bovine Serum Albumin |
| CTCF | Corrected Total Cell Fluorescence |
| FBS | Fetal Bovine Serum |
| GABA | *Gamma*-Aminobutyric Acid |
| GDNF | Glial Cell Line-Derived Neurotrophic Factor |
| HBSS | Hank Balanced Salt Solution |
| NPY | Neuropeptide Y |
| PDGF | Platelet-Derived Growth Factor |
| PYY | Peptide YY |
| RPMI | Roswell Park Memorial Institute |
| RT | Room Temperature |
| Y1R | Y1 Receptor |
| Y2R | Y2 Receptor |

# ABSTRACT

Pain remains a major concern in patients suffering from metastatic cancer to the bone and more knowledge of the condition, as well as novel treatment avenues, are called for. Neuropeptide Y (NPY) is a highly conserved peptide that appears to play a central role in nociceptive signaling in inflammatory and neuropathic pain. However, little is known about the peptide in cancer-induced bone pain. Here, we evaluate the role of spinal NPY in the MRMT-1 rat model of cancer-induced bone pain. Our studies revealed an up-regulation of NPY-immunoreactivity in the dorsal horn of cancer-bearing rats 17 days after inoculation, which could be a compensatory antinociceptive response. Consistent with this interpretation, intrathecal administration of NPY to rats with cancer-induced bone pain caused a reduction in nociceptive behaviors that lasted up to 150 min. This effect was diminished by both Y1 (BIBO3304) and Y2 (BIIE0246) receptor antagonists, indicating that both receptors participate in mediating the antinociceptive effect of NPY. Y1 and Y2 receptor binding in the spinal cord was unchanged in the cancer state as compared to sham-operated rats, consistent with the notion that increased NPY results in a net antinociceptive effect in the MRMT-1 model. In conclusion, the data indicate that NPY is involved in the spinal nociceptive signaling of cancer-induced bone pain and could be a new therapeutic target for patients with this condition.

**Keywords:**

Cancer-induced bone pain; neuropeptide Y; nociception.

# Introduction

Bone metastasis is the main cause of cancer-induced pain (Mercadante, 1997), and derives from some of the most prevalent cancer types, such as breast, prostate and lung cancer (Coleman, 2001, Lozano-Ondoua et al., 2013, Kane et al., 2015). However, managing cancer-induced bone pain has proved to be a challenge, with standardized treatments that lose efficacy over time or present dose-limiting side effects (Mercadante and Portenoy, 2001). Cancer-induced bone pain is characterized by its constant presence (background pain), disrupted by spontaneous and often unpredictable episodes of severe breakthrough pain (Lozano-Ondoua et al., 2013). In spite of its high prevalence and broad impact on the quality of life of cancer patients, the underlying mechanisms inducing and maintaining cancer-induced bone pain are just beginning to be elucidated. Numerous animal models have been developed over the last decades, including the Sprague Dawley MRMT-1 model (Medhurst et al., 2002) that has led to the identification of several spinally expressed peptides, e.g. platelet-derived growth factor (PDGF) (Xu et al., 2016) or glial cell line-derived neurotrophic factor (GDNF) (Meng et al., 2015) involved in metastatic bone pain modulation (Urch et al., 2003, Yang et al., 2014).

Here, we study the role of another spinal peptide transmitter, neuropeptide Y (NPY), a highly conserved 36-amino acid polypeptide present in the central and peripheral nervous systems (Tatemoto, 1982). NPY is abundantly expressed in the rat spinal cord, where up to 15% of the inhibitory interneurons co-express NPY and GABA (Polgár et al., 2010). Both mRNA and NPY-like immunoreactivity (NPY-LI) have been shown to be increased in the lamina I-IV of the dorsal horn of the spinal cord following peripheral nerve injury (Brumovsky et al., 2004). As we have previously reviewed (Diaz-delCastillo et al., 2017), several research groups have shown that NPY affects nociceptive transmission in the spinal cord (Hua et al., 1991, Brumovsky et al., 2007) in animal models of inflammatory (Ji et al., 1994, Taiwo and Taylor, 2002, Intondi et al., 2008, Solway et al., 2011, Taylor et al., 2014), neuropathic (Benoliel et al., 2001, Solway et al., 2011) and postoperative pain (Yalamuri et al., 2013). Moreover, spinal administration of exogenous NPY produces antinociception in inflammatory and neuropathic pain models (Taiwo and Taylor, 2002, Intondi et al., 2008).

The antinociceptive effect of NPY in the spinal cord appears to be exerted by binding to the Y1 receptor (Y1R) (Naveilhan et al., 2001, Taiwo and Taylor, 2002, Shi et al., 2006, Kosti et al., 2013, Malet et al., 2017) and Y2 receptor (Y2R) (Sapunar et al., 2011, Kosti et al., 2013). In the spinal cord, Y1R acts both pre- and post-synaptically (Brumovsky et al., 2002, Brumovsky et al., 2006) and is found in laminae I-V of the dorsal horn (Brumovsky et al., 2006). The Y2R, however, is localized in laminae I or II (Brumovsky et al., 2005, Arcourt et al., 2017), in accordance with its pre-synaptic role on the central terminal of primary afferents (Brumovsky et al., 2005).

In this study we explored the effect of cancer-induced bone pain on the spinal expression of NPY and the availability of NPY binding sites. Moreover, we investigated the antinociceptive potential of intrathecal administration of exogenous NPY in cancer-induced bone pain, and whether this effect is mediated by the Y1R or Y2R.

**Experimental Procedures**

**Study design**

Time-course study: Rats were tested for nociceptive behavior at baseline and days 6, 10, 13 and 16 after surgery, and the disease progression was followed by X-ray and bioluminescent imaging. A subset of cancer-bearing and sham-treated animals was euthanized on postsurgical day 14 (n=10), and another subset on postsurgical day 17 (n=10); the spinal cords were isolated and processed for quantification of NPY-immunoreactivity, and Y1R, Y2R and specific NPY binding.

Pharmacology study: Nociceptive behavior was assessed daily using the limb use and weight-bearing tests. Upon development of nociceptive behavior, pre-defined as a limb use score of 2 and weight-bearing ratio ≤0.35, the animals received a single intrathecal injection of drug or vehicle. The nociceptive behavior (limb use and weight bearing) was then assessed every 30 min, for up to 3 hours. Animals were euthanized and X-ray images of the ipsilateral and contralateral limbs were acquired post-mortem.

All studies were conducted with male and female sham and cancer-bearing rats, and data from both genders were combined, as they showed no significant differences. Cancer-operated animals that did not present bioluminescent signal within 10 days after surgery (indicative of the absence of cancer cells in the tibia) were excluded from the study. All behavioral tests were assessed by the same researcher, blinded to the experimental group.

## Cell Culture

MRMT1-Luc2 cells were cultured in Roswell Park Memorial Institute (RPMI) medium without phenol red or glutamate, supplemented with 10% heat-inactivated Fetal Bovine Serum (FBS) and 1% penicillin-streptomycin-glutamate as previously described (Appel et al., 2017). Prior to surgery, cells were harvested with 0.1% trypsine-EDTA (Invitrogen, Nærum, DK) and re-suspended in Hank Balanced Salt Solution (HBSS) to a working concentration of $5 \times 10^5$ cells/ml. Unless otherwise specified, all reagents were purchased from Thermo Fisher Scientific, DK.

## In vivo

## Animals

All animal experiments were approved by the Danish Animal Experiments Inspectorate (Copenhagen, DK) and performed in accordance with the Danish Act on Animal Experiments (LBK No. 474 of 15/05/2014). Five-week-old male and female Sprague-Dawley rats (Taconic, Tornbjerg, DK) were housed in groups of five same-sex animals in a temperature-controlled room (20-24°C), on a 12-hour light/dark cycle, and provided tap water and feeding (Altromin 1314, Brogaarden, DK) *ad libitum*. A total of 88 animals were included in the studies presented in this manuscript. Animals were housed in standard type IV cages (1805 cm$^2$ of floor space) with Tapvei 2HV bedding. Environmental enrichment was provided as M-bricks (Tapvei, Estonia), nesting material (paper shavings, 2x2 cm), paper ropes and red translucent shelter. Animals were acclimatized to the facility for 1 week and welfare was assessed regularly after the acclimation period and throughout the whole experiment.

## Cancer-induced bone pain surgery

Six-week old rats were anesthetized with isoflurane (3.5% for induction, 2.5% for maintenance; Nomeco, DK) and 5 mg/kg Rimadyl or Norodyl (Pfizer, DK) was administered subcutaneously prior to surgery. The surgery for inoculation of cancer cells was performed as previously described (Falk et al., 2015), modifying the method from Medhurst and associates (Medhurst et al., 2002). Briefly, after shaving and disinfecting the area, an incision was performed on the skin of the medial distal side of the right limb, and the tibia was exposed. A hole was then made in the tibia with a 0.7 mm dental drill, allowing the insertion of a catheter (Smiths Medical, UK) into the proximal medullar cavity. With a 30 G needle, $5 \times 10^3$ MRMT1-Luc2 cells/10 µl HBSS or vehicle (10 µl HBSS) were injected into the catheter. Animals were randomly assigned to sham or cancer group. Upon removal of the catheter, the bone hole was covered with restorative cement (IRM cement, Dentsply, DK), and the wound was closed with two metal clips (11 x 2.5 mm, Agnthos, Sweden). A xylocaine gel (2% w/v; AstraZeneca, Copenhagen, DK) was applied on the surgical site.

**Limb use test**

Rats were allowed to walk freely in a transparent plastic cage with no bedding (500 mm x 300 mm x 500 mm). After 15 min acclimation to the cage in groups of five, the animals were observed individually for 3 min, and their gait was scored as follows: 3=normal use of the cancer-bearing limb, 2=mild or insignificant limping and normal bodyweight distribution, 1=significant limping accompanied by a shift in bodyweight distribution towards the healthy limb, 0=partial or total lack of use of the cancer-bearing limb (shown as momentarily holding the limb aloft while in locomotion and/or sitting), as previously described (Appel et al., 2017). A humane endpoint was established for a limb use score of 0.

**Weight-bearing test**

The animals were placed in incapacitance tester (MJS Technology Ltd., Buntingford, Herfordshire, UK) consisting of two separate scales where the individual weight placed on each limb can be measured. The test was performed for 4 seconds, and three measurements were obtained per animal. The weight-bearing ratio was calculated as the average weight placed on the cancer-bearing limb divided by the total weight placed on both limbs.

## X-ray imaging

X-ray images of anaesthetized rats (3.5% isoflurane for induction, 2.5% for maintenance, Nomeco, DK) were captured in a Lumina XR Apparatus (Caliper Life Science, Teralfene, Belgium). Every image was calibrated to a standard aluminum wedge. Images were analyzed with Image J (National Institute of Health, USA) by subtracting the mean gray scale value of two background soft-tissue regions, from the gray scale value of the distal area of the tibia. All measurements were normalized against a standard aluminum wedge. Data analyses were performed by a researcher blinded to the experimental groups.

## Bioluminescence imaging

D-Luciferine (Caliper Life Sciences, Teralfene, Belgium) was dissolved in PBS and administered to cancer-operated animals by intraperitoneal injection (40 mg/kg), 10 min before bioluminescence imaging was carried out. The rats were placed in a Lumina XR Apparatus (Caliper Life Sciences, Teralfene, Belgium) while anaesthetized with isoflurane (2.5-3.0%, Nomeco, DK). Three images were captured for each animal, with an exposure time of 5-120 seconds, binning M (4) and F/stop 1. The bioluminescent signal was measured with the IVIS Imaging Software (Living Image©, Caliper Life Sciences, Teralfene, Belgium) by adjusting the threshold to 35% of the signal for each image. The average signal of three pictures per animal is reported, and the readout is measured in total flux (photons/s).

## Drug administration

Upon development of nociception, rats were anaesthetized with 2.5% isoflurane (Numeco, DK) and randomly selected for intrathecal injection with drug or vehicle treatment. Treatments were: 1) either isotonic saline vehicle (n=6), NPY (n=6), BIBO3304 (Y1R antagonist; n=6) or NPY with BIBO3304 (n=7), or 2) either 10% DMSO vehicle (n=5), NPY dissolved in 10% DMSO (n=7), BIIE0246 (Y2R antagonist) dissolved in 10% DMSO (n=5) or BIIE0246 with NPY dissolved in 10% DMSO (n=6). The doses of the drugs were 30 µg (≈ 7 nmol) NPY, 3 µg (≈ 4 nmol) BIBO3304 and 3 µg (≈ 3 nmol) BIIE0246, and the injected volume was 10 µl in all cases. The NPY and antagonists doses were selected in accordance with previous studies, which showed a dose-dependent effect (Taiwo and Taylor, 2002, Smith et al., 2007, Intondi et al., 2008). Additionally, 30 µg NPY has been shown to not cause nonspecific motor effects in Sprague-Dawley rats (Taiwo and Taylor, 2002). Successful placement of the needle was confirmed by flitching of the tail.

**Materials**

Human NPY with C-terminal amidation (Schafer-N, Copenhagen, DK) was prepared as 30 µg/10 µl in isotonic saline solution with 0.01% bovine serum albumin (BSA; Sigma Aldrich, DK) or DMSO solution (0.01% BSA and 10% DMSO) before use. The Y1R antagonist BIBO3304 (#2412, Tocris Bioscience, Bristol, UK) was diluted to a concentration of 3 µg/10 µl in isotonic saline solution with 0.01% BSA before use. The Y2R antagonist BIIE0246 (#1700, Tocris Bioscience, Bristol, UK) was diluted to a concentration of 3 µg/10 µl DMSO solution before use.

**Tissue processing**

Animals were deeply anaesthetized with 3.5% isoflurane (Nomeco, DK) and decapitated. The lumbar region of the spinal cord, including L3 to L6, was extracted and freshly frozen on dry ice or liquid nitrogen. The tissue was then sectioned using a cryostat into 14-µm thick slices (CM3050S, Leica, DK).

**Immunohistochemistry**

Sections were washed 2x5 min with K-PBS (150 mM NaCl, 2.7 mM KCl in 0.05% PBS) and fixated with 4% paraformaldehyde for 20 min. After 3x5 min washing in T-KPBS (0.1% Triton X-100 in KPBS), 3x5 min washing in K-PBS and 1 hour incubation with 1% BSA (A9647-50G, Sigma-Aldrich, USA) in 0.3 M glycine T-KPBS at room temperature (RT), sections were incubated at 4°C over night with 1:10,000 rabbit primary antibody against NPY (N9528, Sigma-Aldrich, DK). Next, slides were washed 3x5 min in K-PBS and incubated 2 hours at RT with 1:1000 secondary anti-rabbit antibody Alexa Fluor 488 (A3631, Invitrogen, DK). Finally, following 3x5 min washing with K-PBS, the slides were mounted with Fluoroshield Mounting Medium with DAPI (ab104139, Abcam, DK) and imaged in a Zeiss Axioscop2 (Carl Zeiss Microscopy, LLC, USA) microscope. Pictures were taken using the Zeiss Axio Imager Z1 software (AxioVision 4.9.1, Carl Zeiss Microscopy GmbH) and analyzed with Image J. As previously described (McCloy et al., 2014), the CTCF (corrected total cell fluorescence) was calculated as the integrated density of the stained dorsal horn area, minus the product of the background fluorescence times the area of the stained dorsal horn. Background fluorescence was calculated as the mean integrated density of three random, non-stained areas. A minimum of four pictures was used to determine the average CTCF. Percentage of fluorescence increase was calculated as the difference between ipsilateral and contralateral CTCF, divided by the contralateral CTCF. Data analyses were performed by a researcher blinded to the experimental groups.

## NPY binding autoradiography

NPY receptor autoradiography was carried out as previously described (Woldbye et al., 2005, Elbrønd-Bek et al., 2014). Procedures were performed at RT if not stated otherwise. Spinal cord sections were defrosted for 30 min before pre-incubation for 20 min in binding buffer (25 mM HEPES, 2.5 mM $CaCl_2$, 1 mM $MgCl_2$, 0.5 g/L bacitracin, and 0.5 g/L BSA, pH 7.4). Then, sections were incubated for 60 min in binding buffer containing 0.1 nM [125I][Tyr36]-mono-iodo-peptide YY ([125I]-PYY; 4000 Ci/mmol, porcine synthetic, #NEX2400, PerkinElmer, DK) alone to visualize total binding to all NPY receptors, or with addition of 1 µM unlabeled NPY to determine nonspecific binding. For visualization of Y1R or Y2R binding, binding buffer with 0.1 nM [125I][Tyr36]mono-iodo-PYY was used with addition of 10 nM PYY3-36 (Y2 preferring agonist; mouse synthetic, #H-8585, Bachem AG) or with 10 nM Leu31,Pro34-NPY (Y1 preferring agonist; mouse synthetic, #H-3306, Bachem AG), respectively. These incubations were terminated by brief rinse and subsequent wash in binding buffer for 2x30 min. Sections were air-dried and exposed to 125I-sensitive Kodak Biomax MS films with 125I-microscales (Amersham Bioscience, DK) for 4 days, and films were then developed in Kodak D19 developer. Levels of total NPY binding, Y1R and Y2R binding, nonspecific binding were quantified using Image J by measuring the average optical density on four sections from L3 to L6 (Watson et al., 2009) from each spinal cord. Film background was subtracted from each measurement. Specific receptor binding was calculated by subtracting the nonspecific binding. Data analyses were performed by a researcher blinded to the experimental groups.

**Statistical procedures**

All plots were generated in GraphPad Prism 7.0 (Graph Pad Inc, CA, USA). All behavioral data were analyzed by one-way ANOVA or repeated measures two-way ANOVA; statistical significance was determined using the Bonferroni correction for multiple comparisons. Immunohistochemistry results and binding assays were analyzed by unpaired, two-tailed, Student's t-tests. All data are presented as mean ± standard error of the mean (S.E.M.). The level of significance for all tests was set at p <0.05.

## 2 Results

**Progression of the nociceptive behavior**

The MRMT1-Luc2 rat model of CIBP was used to study the effect of bone cancer on the NPY system over time. Compared to sham animals, a significant decrease in the limb use score was observed in cancer-bearing rats (p<0.0001 at days 10, 13 and 16; n=5-10) **(Fig. 1A)**. Sham animals were unaffected and maintained a constant limb use score of 3 throughout the experiment.

Correspondingly, cancer-bearing animals presented a significant decrease in weight-bearing ratio from day 10 after surgery, compared with sham (p<0.01 at day 10, p<0.0001 at day 13, and 16, n=5-10). In contrast, sham animals maintained a constant weight-bearing ratio throughout the whole experiment **(Fig. 1B)**.

**Progression of the cancer-induced bone loss**

To follow the cancer-induced bone degradation, the relative bone mass of sham and cancer-bearing animals was assessed with X-ray densitometry **(Fig. 1C, E)**. Compared with sham, a significant decrease in the relative bone density of cancer-bearing rats was observed from post-surgical day 10 (p<0.0001, n=5-10). Sham rats showed no significant variation in their relative bone density over time **(Fig. 1C, E).** Tumor burden was assessed with bioluminescence on 6, 10, 13 and 16 days after surgery. This signal was present at day 6 after surgery and increased over time, reaching its maximum at postsurgical day 13 **(Fig. 1D, E)**.

### NPY is up-regulated in the spinal cord of cancer-bearing rats

To study the effect of cancer-induced bone pain on NPY expression in the spinal cord, NPY-like immunoreactivity (NPY-LI) was measured in the ipsilateral and contralateral dorsal horn of sham and cancer-bearing rats 14 and 17 days after surgery **(Fig. 2A-C)**. The percentage change in NPY-LI in the ipsilateral over the contralateral dorsal horn of cancer-bearing rats was significantly higher 17 days after surgery compared with sham animals (p<0.05, n=10). No changes were found 14 days after surgery (n=7-10) **(Fig. 2A)**.

### NPY binding capacity is preserved in the spinal cord of cancer-bearing animals

To elucidate specific NPY receptor binding capacity in cancer-induced bone pain, radiolabelled [(125)I]-peptide Y (PYY) was used to study total NPY receptor binding **(Fig 3A, D)**. A mixture of PYY3-36, a Y2R agonist, and radiolabelled [(125)I]-PYY was used to visualize the receptor binding profile of Y1R **(Fig 3B, D)**, while (Leu31, Pro34)-NPY, a Y1 agonist, was used in combination with radiolabelled [(125)I]-PYY to study the binding profile of the Y2R **(Fig 3C, D)**. No significant changes were found neither in specific total binding, Y1 and Y2 binding nor in non-specific binding between the ipsi- and contralateral dorsal horn of sham and cancer-bearing rats 17 days after surgery, indicating that the number of available NPY receptors is not altered in the cancer-induced bone pain model (n=10) **(Fig 3A-D).**

### NPY induces antinociception in CIBP through both Y1R and Y2R

To test the antinociceptive potential of exogenous NPY, cancer-bearing rats presenting nociceptive behavior were administered an intrathecal injection of 30 µg NPY or vehicle, and their limb use and weight-bearing ratio were tested every 30 minutes thereafter.

Intrathecal administration of NPY induced a significant increase in the limb use score over the course of 3 hours compared to vehicle (NPY in saline: p<0.001, n=6-7; NPY in 10% DMSO: p<0.005, n=5-7) **(Fig 4A, C)**. Consistent with this finding, there was a significant improvement in the weight-bearing ratio of animals receiving NPY in saline (p<0.01 at 60, 90 and 120 min, p<0.05 at 150 min, n=6-7) that lasted up to 2.5 hours when compared to baseline **(Fig. 4B)**. Similarly, NPY dissolved in 10% DMSO caused improvement in weight-bearing, yet significant effect was only reached 90 and 150 min after administration (p<0.01 at 90 min, p<0.05 at 150 min, n=5-7) **(Fig. 4D)**.

Next, we studied whether NPY exerts its antinociceptive effect through the Y1R or Y2R. The Y1R antagonist, BIO3304, abolished the antinociceptive effect of NPY **(Fig. 4A, 4B),** while intrathecal delivery of BIBO3304 alone did not affect neither the limb use score nor the weight-bearing ratio of cancer-bearing rats. Likewise, the antinociceptive effect of NPY dissolved in 10% DMSO was abolished by the Y2R antagonist BIIE0246 **(Fig. 4C, 4D)**. Intrathecal application of BIIE0246 alone also did not affect the limb use score or weight-bearing ratio of cancer-bearing rats.

## Discussion

This study demonstrates an antinociceptive effect of spinally administered NPY in a cancer-induced bone pain model, mediated through both Y1R and Y2R. The study also shows an increased NPY-LI and unaltered specific binding to the Y1R and Y2R in the spinal cord of cancer bearing animals, suggesting NPY as a potential target in cancer-induced bone pain.

Even though the mechanisms underlying NPY-mediated antinociception are still not fully understood, it has been hypothesized that NPY acts as a homeostatic mediator, since its increased expression in the dorsal root ganglia and spinal cord after injury may promote blockage of the nociceptive signal to supraspinal regions (Brumovsky et al., 2004). The inhibitory role of endogenous NPY in the transition from acute to chronic pain has been proposed in animal models of inflammatory and neuropathic pain (Solway et al., 2011, Taylor et al., 2014), where NPY expression is increased in the dorsal horn of the spinal cord by nociceptive stimuli (Ji et al., 1994, Brumovsky et al., 2004). In agreement with this hypothesis, we find that NPY is significantly increased in the MRMT1-Luc2 rat model of metastatic bone pain 17 days after surgery, when the disease is more advanced than 14 days after surgery. However, it should be noted that there is a possibility of extravasation of cancer cells from the cancer-bearing bone into the adjacent soft tissue or nerve fibers, which could lead to the development of extraskeletal nociception influencing the spinal upregulation of NPY. Additionally, the present study does not address whether the increase in NPY in the dorsal horn has a spinal or ganglionic origin; previous studies report that novel NPY synthetized in the DRGs can be transported towards the axon terminals of spinal sensory neurons (Ohara et al., 1994, Ossipov et al., 2002). Moreover, it has been suggested that in the sciatic nerve injury model, the increase in spinal NPY has both a spinal (when NPY-LI is located in laminae I and II) and ganglionic origin (when NPY-LI is located in laminae III and IV) (Ohara et al., 1994). Further studies could contribute to elucidating whether the MRMT-1 model of cancer-induced bone pain presents a similar pattern.

The specific binding density of [(125)I]-PYY, a radiolabelled NPY analog, was similar between sham and cancer-bearing animals 17 days after surgery. Likewise, the binding levels of Y1R and Y2R in sham and cancer-bearing rats were also similar. Similarly, no differences were found in the amount of spinal Y1R-LI of sham rats and those undergoing "medium" or "light" single ligature nerve constriction (SLNC) (Brumovsky et al., 2004). Nonetheless, decreased Y1R-LI was found on the dorsal horn of rats undergoing "strong" SLNC or axotomy, highlighting the dynamic role of the NPY system in different pain models. Considering the observed increase in spinal NPY-LI under conditions of unaltered NPY receptor binding levels, we suggest that there may be increased net NPY signaling in the cancer-induced bone pain condition at 17 days after surgery. This is further consistent with the concept that increased NPY synthesis in this model could be a compensatory antinociceptive response, as has been suggested in other pain models (Wakisaka et al., 1991, Solway et al., 2011). Importantly, changes in the spinal expression of NPY seem to depend on the particular nociceptive input, animal model or time of euthanasia. While previous studies report an upregulation of NPY-LI in rat models of neuropathic (Brumovsky et al., 2004) and inflammatory pain (Ji et al., 1994), Honore et al. found a comparable NPY upregulation in a mouse model of neuropathic pain, but reported no differences in spinal NPY expression in a mouse model of inflammatory or sarcoma-induced bone pain (Honore et al., 2000), stressing the distinct role of the NPY system across species and animal models.

To further address if NPY signaling could exert an antinociceptive effect in cancer-induced bone pain, we studied the effects of spinal administration of NPY in the early phase of nociception development of the MRMT-1 model. We found that NPY exerted an antinociceptive effect that lasted up to three hours, similarly to its effect in several models of inflammatory, neuropathic and postsurgical pain (Taiwo and Taylor, 2002, Intondi et al., 2008, Yalamuri et al., 2013, Taylor et al., 2014). This analgesic effect seemed less potent in the cancer-induced bone pain model compared with other animal models of inflammatory or neuropathic pain, where similar doses of NPY have been found to restore baseline behavioral responses (Intondi et al., 2008). Cancer-induced bone pain, however, is a complex pain state that comprises both inflammatory and neuropathic mechanisms, but presents a unique set of neurochemical changes as well (Luger et al., 2002, Falk S, 2014). Therefore, providing analgesia to animal models of bone cancer pain is often challenging; accordingly, previous studies have shown a need for higher opioid doses for the treatment of bone cancer pain than other pain conditions, such as inflammatory pain (Luger et al., 2002).

To test whether the Y1R and the Y2R mediate the NPY-induced antinociceptive effect, we administered NPY together with an antagonist for each of the receptors. While both NPY and the BIBO3304 can be dissolved in saline, the low solubility of BIIE0246 required addition of 10% DMSO, an approach previously used by other research groups (Intondi et al., 2008). We therefore tested intrathecal administration of NPY in both saline and a 10% DMSO vehicle, and found similar effects in the limb use and weight-bearing tests.

Both Y1R and Y2R are present in the dorsal horn of the spinal cord, where Y1R is highly expressed in laminae I-V (Brumovsky et al., 2006) and Y2R is specially abundant in lamina II (Arcourt et al., 2017). Several studies have confirmed the involvement of the Y1R in NPY-mediated analgesia, including both knock-out studies (Naveilhan et al., 2001) and the use of saporin-conjugated inhibition of Y1R$^+$ neurons (Wiley et al., 2009). While the role of Y2R has been established in models of acute (Hua et al., 1991) and chronic neuropathic pain (Xu et al., 1999, Solway et al., 2011), its involvement in inflammatory pain remains controversial (Taiwo and Taylor, 2002). Our results suggest that each receptor is necessary for the NPY-mediated anti-nociception in cancer-induced bone pain, since both the Y1 (BIBO3304) and the Y2 (BIIE0246) receptor antagonists completely abolished the analgesic effect of intrathecal NPY.

Previous studies in mouse models of inflammatory or neuropathic pain have shown that the administration of either a Y1R antagonist or a Y2R antagonist alone worsens tactile and thermal hypersensitivity, probably by inhibiting the analgesic effect of endogenous NPY (Solway et al., 2011). However, in this study of CIBP, the administration of either of the antagonists alone did not elicit changes in nociception. The antagonists were given at a time point at which the immunohistochemistry showed no apparent increase in NPY in the spinal cord dorsal horn of rats suffering from CIBP, indicating that it was unlikely that endogenous NPY elicited an analgesic effect. Therefore, the Y1R and Y2R antagonists would not be expected to affect the pain-related behaviors. Yet, our data suggest that a single injection of exogenous NPY is enough to temporarily reduce nociception in cancer-induced bone pain.

We suggest that the antinociceptive action of NPY in metastatic bone pain may be mediated by both inhibition of Y1R-positive excitatory interneurons (Brumovsky et al., 2006), thereby suppressing the supraspinal transmission of the nociceptive signal, and inhibition of the release of excitatory neurotransmitters from primary afferents through its action on the Y2R (Brumovsky et al., 2005). The dual action of spinal NPY could explain why both BIBO3304 and BIIE0246 abolished the anti-nociceptive effect.

NPY has been previously acknowledged as a promising target for the treatment of inflammatory and neuropathic pain. Taken together, our studies suggest that NPY and its receptors Y1 and Y2 might also be a promising target for the development of future treatments for CIBP.

## Conflict of interest

The authors present no conflict of interest to disclose.

## Author contributions

M.D., S.F., S.H., D.W, A.H. designed the experiments. M.D., C.A., S.F., S.H. performed the experiments and analyzed the data. M.D.C., S.F., C.K.A., A.H., D.W. wrote the manuscript. All authors critically revised and accepted the manuscript.

## Acknowledgements

We would like to thank Camilla Skåstrøm Dall, Tina Maria Estrup Axen and Mathilde Caldara for their technical assistance. All authors of this manuscript have contributed to data collection, interpretation and writing of this manuscript. This project has received funding from the European Union's Horizon 2020 research and innovation program under the Marie Skłodowska-Curie grant agreement No 642720.

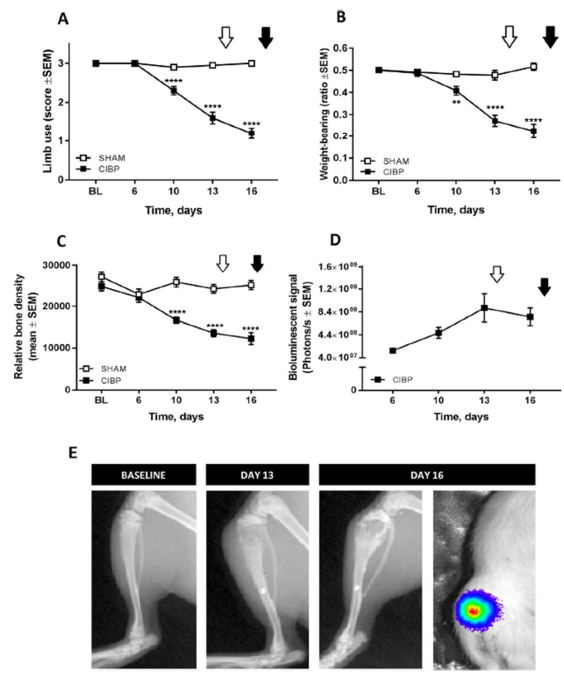

**Figure 1.** Cancer progression leads to the development of nociceptive behavior and a decrease in relative bone mass over time. **A.** Cancer-bearing rats showed a significant decrease in limb use score from day 10 after surgery compared to sham **B.** A significant decrease in the weight-bearing ratio of cancer-bearing animals was also found from day 10. **C.** Cancer-bearing rats presented a significant decrease in relative bone density from post-surgical day 10. **p $<0.01$, ****p $<0.0001$; n=5-10. Two-way ANOVA followed by Bonferroni post hoc test. **D.** All cancer-bearing rats showed a detectable bioluminescence signal from 6 days after surgery, indicating the presence of living tumor cells. White arrows indicate time of euthanasia for the first group of animals, 14 days after cancer inoculation, and black arrows indicate time of euthanasia for the second group of cancer-bearing animals, 17 days after cancer inoculation. **E.** Representative X-ray images and bioluminescent signal of a cancer-bearing rat.

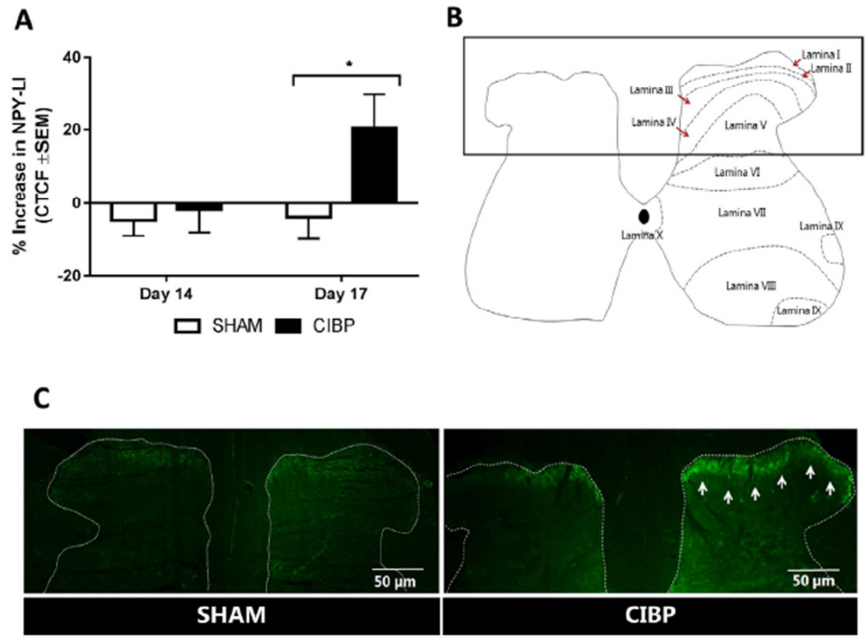

**Figure 2.** Spinal NPY is upregulated in the late stage of cancer-induced bone pain. **A.** NPY-LI was significantly increased in the ipsilateral over the contralateral dorsal horn of cancer-bearing rats, but not sham rats, 17 days after surgery (*p <0.05; n=10, Student´s t-test). There was no difference in NPY-LI 14 days after surgery (n=7-10). **B.** Sketch of the spinal cord; NPY-LI was measured in the dorsal horn of the spinal cord, approximately laminae I-II. **C.** Representative pictures of NPY-LI a spinal cord section of a sham and a cancer-bearing rat euthanized 17 days after surgery; note the increased staining in the ipsilateral dorsal horn of the cancer-bearing rat, indicated with arrows.

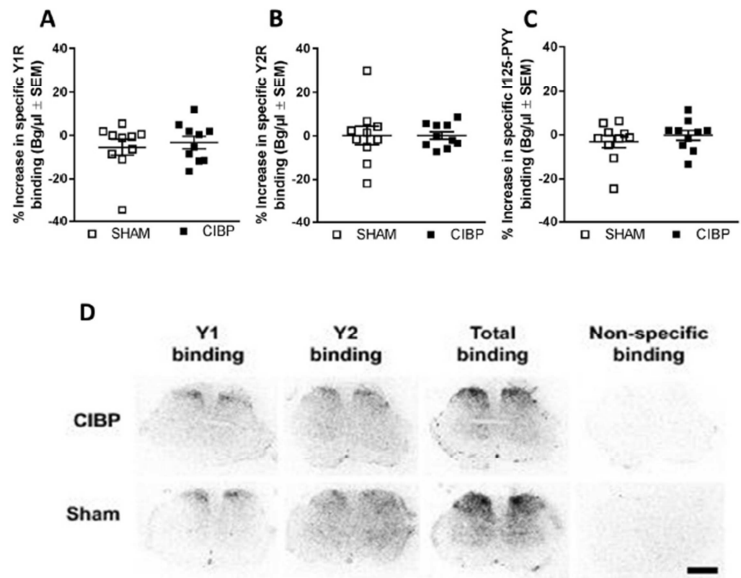

**Figure 3.** NPY binding sites are conserved in the spinal cord of cancer-bearing rats. **A.** There were no significant differences in the specific binding of Y1R between sham and cancer-bearing rats (n=10). **B.** The specific binding of the Y2R did not differ between cancer-bearing and sham rats (n=10). **C.** The specific total NPY binding on the ipsilateral over the contralateral dorsal horn was similar between cancer-bearing and sham rats (n=10). **D.** Representative picture of the Y1, Y2, total and non-specific binding assays. *Magnification bar=1mm.*

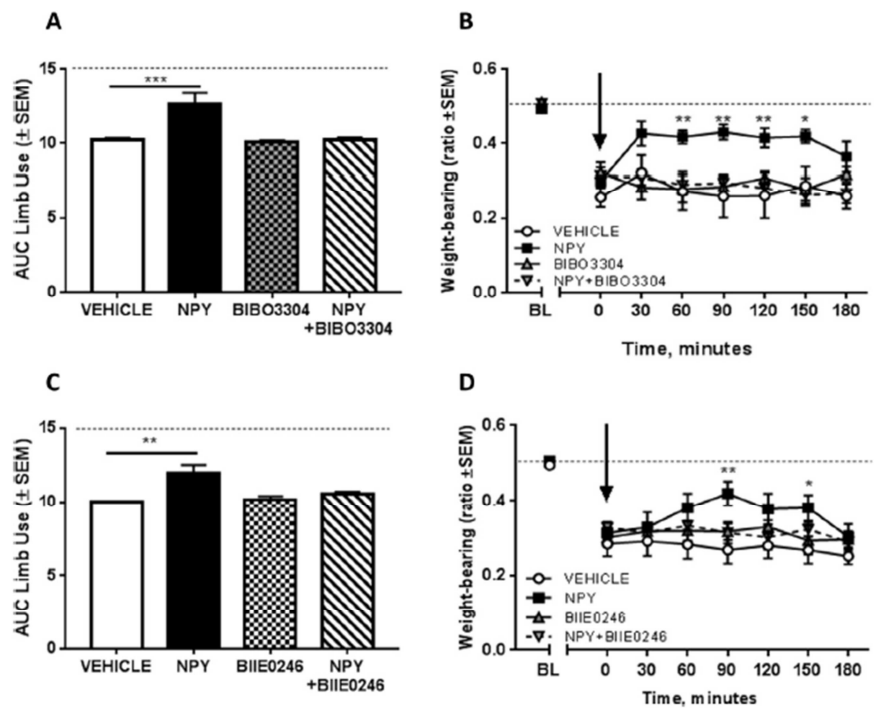

**Figure 4.** Intrathecal NPY administration induces analgesia in cancer-induced bone pain, through its action on the Y1 and Y2 receptors. **A, B.** Spinal administration of NPY (at time 0, indicated with an arrow in **B**) induced a significant increase in the limb use (**A**, shown as area under the curve) and weight-bearing ratio (**B**) of cancer-bearing rats, compared to vehicle. This effect lasted up to 2.5 hours, and it was abolished by the Y1R antagonist BIBO3304. Administration of BIBO3304 alone did not increase nociception. **C, D**. Intrathecal administration of NPY (at time 0, indicated with an arrow in **D**) induced an analgesic effect shown in the limb use (**C**, shown as area under the curve) and weight-bearing ratio (**D**), compared to vehicle. This effect was reversed by the Y2R antagonist BIIE0246. Dashed lines indicate the baseline values of naïve animals. *p <0.05, **p <0.01, ***p <0.001; n=5-7. Bonferroni-adjusted test following significant one or two-way ANOVA.